\documentclass[12pt]{iopart}

\usepackage{amsfonts}
\usepackage{amssymb}
\usepackage{array}
\usepackage{floatflt}
\usepackage{graphicx}
\newcommand*{\eq}[1]{(\ref{#1})}

\begin{document}
\title[Entanglement of systems of nuclear spins at the adiabatic demagnetization]{Entanglement of systems
of dipolar coupled nuclear spins at the adiabatic demagnetization}
\author{S. I. Doronin$^1$, E.~B.~Fel'dman$^1$, M. M. Kucherov$^2$, A.~N.~Pyrkov$^1$}
\address{$^1$Institute of Problems of Chemical Physics of Russian Academy of Sciences, Chernogolovka, Moscow Region, Russia, 142432}
\address{$^2$Siberian Federal University, Krasnoyarsk, 660074, Russia}
\ead{pyrkov@icp.ac.ru}

\begin{abstract} We consider the adiabatic demagnetization in the rotating reference frame (ADRF) of a
system of dipolar coupled nuclear spins $s=1/2$ in the external magnetic field. The demagnetization starts with the offset of the external
magnetic field (in frequency units) from the Larmor frequency being several times greater than the local dipolar field.
For different subsystem sizes, we have found from numerical simulations the temperatures at which subsystems of a one-dimensional
nine-spin chain and a plane nine-spin cluster become entangled. These temperatures are of the order of microkelvins and are almost independent
of the subsystem size. There is a weak dependence of the temperature on the space dimension of the system.
\end{abstract}

\date{\today}
\maketitle

\section{Introduction}
Entangled states are very important for quantum computing, teleportation, cryptography \cite{chuang}. Although
entanglement is a profound concept of quantum information theory its experimental realization in many-body systems is an
unsolved problem up to now. Meanwhile the existing criterions of the existence of entanglement \cite{wootters,peres} allow us to
investigate entangled states in conventional NMR experiments \cite{goldman}. It is well known that entanglement emerges in systems of nuclear
spins at microkelvin temperatures \cite{pyr,jetp}. Such temperatures can be achieved with the adiabatic demagnetization in the rotating
reference frame (ADRF) \cite{abraham}.

In the present article we consider the emergence of entanglement in
the course of the ADRF in the system of nuclear spins coupled by the
dipole-dipole interaction (DDI). We perform numerical calculations
for a chain consisting of nine spins and a plane nine-spin cluster
in order to prove the emergence of entanglement of its different
subsystems. We use the Wootters criterion~\cite{wootters} for the
investigation of the spin pair entanglement. Entanglement of
subsystems with larger numbers of spins is investigated with the
positive partition transposition (PPT) criterion~\cite{peres}.
Entanglement emerges at approximately the same temperature of the
order of microkelvins for all subsystem sizes. However, the
temperature depends on the space dimension of the system.

\section{The density matrix of a spin system at the ADRF}
We consider a system of $N$ nuclear spins $s=1/2$ coupled with the external magnetic field. The system is irradiated by the
frequency-modulated radio-frequency (rf) field, $w(t),$ which is perpendicular to the permanent magnetic field. The Hamiltonian, $H_{lab},$ of
the system in the laboratory frame can be written as follows 
\begin{equation} H_{lab}=w_0I_z+H_{dz}+2w_1I_x\cos\left[\int_0^tw(t')dt'\right],
\end{equation}
where $w_0$ is the Larmor frequency, $w_1$ is the amplitude of the
rf field (in frequency units), $I_{n\alpha}$ is the projection of
the angular momentum operator of the n-th spin ($n=1,2\ldots N$) on
the $\alpha$ axis ($\alpha=x,y,z$),
$I_\alpha=\sum_{n=1}^NI_{n\alpha},$ and $H_{dz}$ is the secular part
of the DDI Hamiltonian \cite{goldman} which can be written as
\begin{equation}
H_{dz}=\sum_{i<j}D_{ij}(3I_{iz}I_{jz}-\overrightarrow{I_i}\overrightarrow{I_j}),
\end{equation}
where $D_{ij}$ is the DDI coupling constant of spins $i$ and $j,$
and
$\overrightarrow{I_i}\overrightarrow{I_j}=I_{ix}I_{jx}+I_{iy}I_{jy}+I_{iz}I_{jz}.$
The spin dynamics of the system is determined by the density matrix,
$\rho(t),$ whose time evolution occurs according to the Liouville
equation ($\hbar=1$)~\cite{goldman} \begin{equation}
i\frac{d\rho(t)}{dt}=[H_{lab},\rho(t)].
\end{equation}
Converting the density matrix, $\rho(t),$ with the unitary
transformation \begin{equation} \rho(t)=e^{-i
I_z\int_0^tw(t')dt'}\rho^*(t)e^{i I_z\int_0^tw(t')dt'}
\end{equation} and neglecting the irrelevant terms oscillating with
the double Larmor frequency one obtains the evolution equation for
the density matrix $\rho^*(t)$ \begin{equation} \label{liuv}
i\frac{d\rho^*(t)}{dt}=[(w_0-w(t))I_z+H_{dz}+w_1I_x,\rho^*(t)].
\end{equation}
According to \eq{liuv} the Hamiltonian, $H,$ of the system can be
written in the rotating reference frame (RRF) as \begin{equation}
H=\Delta(t)I_z+H_{dz}+w_1I_x, \end{equation} where
$\Delta(t)=w_0-w(t)$ is the resonance offset of the longitudinal
frequency from the Larmor frequency. In the course of the ADRF, the
offset, $\Delta(t),$ changes slowly in order to satisfy to the
adiabatic condition \cite{goldman} \begin{equation}
\frac{|\dot{\Delta}(t)|}{\pi w_1^2}\ll1 \label{adiabat_cond}.
\end{equation} The condition~\eq{adiabat_cond} means that the
resonance offset, $\Delta(t),$ changes so slowly that the spin
system is in the quasi-equilibrium state at every moment of
time~\cite{goldman}. Thus the spin system can be described by the
thermodynamic quasi-equilibrium density matrix, $\rho_{eq}(t),$ as
\begin{equation} \label{rho} \rho_{eq}(t)=e^{-\beta H}/Z,
\end{equation} where $\beta$ is proportional to the inverse
temperature ($\beta=\hbar/kT$) and $Z$ is the partition function. It
is important to emphasize that the temperature, $T,$ is the spin
temperature of the system which is isolated from all other degrees
of freedom. The reason of such isolation is long spin-lattice
relaxation times which exceed spin-spin relaxation times by several
orders of magnitude~\cite{goldman}. The important consequence for
our approach is the following. The decoherence effects at
entanglement generation with ADRF are irrelevant.

The entropy, $S,$ of the system is given by \cite{goldman}
\begin{equation} \label{eentr}
S=-k\tr\{\rho_{eq}(t)\ln[\rho_{eq}(t)]\}.
\end{equation}
We start the ADRF with $\Delta\gg w_{loc}$ where
$w_{loc}=\{\tr{\{H_{dz}^2\}}/\tr{[I_z^2]}\}^{1/2}$ is the local
dipolar field. Then the offset, $\Delta(t),$ is the linear function
of time \begin{equation} \Delta=\Delta_0-at \end{equation} where
$\Delta_0$ and $a$ are the given quantities. Since the entropy,
$S=const$, one can obtain the inverse temperature, $\beta(t),$ in
the course of the ADRF, if the offset, $\Delta(t),$ is known. It
means that we obtain the density matrix, $\rho_{eq}(t),$ during the
ADRF. In particular, the entropy, $S,$ is \begin{equation}
\label{entr}
S=kN\ln2+kN\ln[\cosh\left(\frac{\beta\Delta}2\right)]-\frac{k}2N\beta\Delta\tanh\left(\frac{\beta\Delta}2\right)
\end{equation}
at $\Delta\gg w_{loc}$. \eq{entr} can be used in order to find the
initial inverse temperature of the system. \section{The reduced
density matrix of a spin pair at the ADRF and the Wootters
criterion} In order to obtain the reduced density matrix of an
arbitrary pair of spins $i$ and $j$ we use the approach developed in
\cite{pyr,jetp}. The density matrix, $\rho_{eq}(t),$ of \eq{rho} can
be represented as \cite{pyr,jetp} \begin{equation} \label{basmatr}
\rho=\sum_{\xi_1,\xi_2,\ldots,\xi_N=0}^3\alpha_{12\ldots
N}^{\xi_1\xi_2\ldots\xi_N}x_1^{\xi_1}\otimes\ldots\otimes
x_N^{\xi_N},
\end{equation}
where $N$ is a number of spins in the system, $\xi_k \
(k=1,2,\ldots,N)$ is one of the values $\{0,1,2,3\},$ $x_k^0=I_k$ is
the unit matrix of the dimension $2\times 2,$ $x_k^1=I_{kx},$
$x_k^2=I_{ky},$  $x_k^3=I_{kz},$ and $\alpha_{12\ldots
N}^{\xi_1\xi_2\ldots\xi_N}$ is a numerical coefficient. Averaging
the density matrix of \eq{basmatr} over all spins except spins $i$
and $j$ and taking into account that $\tr\{x_k^{\xi_k}\}=0\
(k=1,2,\ldots,N;\,\xi_k=1,2,3)$ we arrive at the following
expression for the reduced density matrix, $\rho_{eq}^{(ij)}(t),$ of
the $i$th and $j$th spins \begin{equation} \label{rmatr}
\rho_{eq}^{(ij)}(t)=\sum_{\xi_i,\xi_j=0}^3\alpha_{ij}^{\xi_i\xi_j}x_i^{\xi_i}\otimes
x_j^{\xi_j},
\end{equation}
where
\begin{equation}
\label{alpha} \alpha_{ij}^{\xi_i\xi_j}=\frac{2^{N-2}\tr\{\rho
x_i^{\xi_i}x_j^{\xi_j}\}}{\tr\{(x_i^{\xi_i})^2(x_j^{\xi_j})^2\}}.
\end{equation}
The coefficients, $\alpha_{ij}^{\xi_i\xi_j},$ of \eq{alpha} can be
calculated numerically. Then the reduced density matrix of the pair
of spins $i$ and $j$ is determined completely. In order to apply the
Wootters criterion \cite{wootters} one should find the
"spin-flipped" density matrix, $\tilde{\rho}_{eq}^{(ij)}(t),$ which
is \begin{equation} \label{flip}
\tilde{\rho}_{eq}^{(ij)}(t)=(\sigma_y\otimes\sigma_y)[\rho_{eq}^{(ij)}(t)]^*(\sigma_y\otimes\sigma_y)
\end{equation}
where the asterisk denotes complex conjugation in the standard basis
$\{|00\rangle,|01\rangle,|10\rangle,|11\rangle\}$ and the Pauli
matrix $
\sigma_y=\left(\begin{array}{cc}0&-i\\i&0\end{array}\right). $ The
calculation of the density matrix, $\tilde{\rho}_{eq}^{(ij)}(t),$
and the diagonalization of the matrix product
$\rho_{eq}^{(ij)}(t)\tilde{\rho}_{eq}^{(ij)}(t)$ are performed
numerically. The concurrence of the two--spin system with the
density matrix $\rho_{eq}^{(ij)}(t)$ is equal to~\cite{wootters}
\begin{equation} C=max\{0,
2\lambda-\lambda_1-\lambda_2-\lambda_3-\lambda_4\},\quad
\lambda=max\{\lambda_1,\lambda_2, \lambda_3, \lambda_4\}
\end{equation} where $\lambda_1,$ $\lambda_2,$ $\lambda_3,$ and
$\lambda_4$ are the square roots of the eigenvalues of the product
$\rho_{eq}^{(ij)}(t)\tilde{\rho}_{eq}^{(ij)}(t).$
\section{The entangled state of a spin subsystem with its environment at the ADRF} The spin pair
entangled states are a simple type of entanglement which can be described with the Wootters criterion \cite{wootters} completely. Meanwhile entanglement of bigger subsystems of the system can also emerge in the course of the ADRF. A necessary condition of separability is described by the Peres criterion~\cite{peres} which is known as the positive partition transposition (PPT) criterion~\cite{peres}. According to the criterion, entanglement emerges, if after the transposition of the density matrix over variables of one subsystem there are some negative eigenvalues of the density matrix of the system. We wrote a program which allows us to investigate entanglement of an arbitrary subsystem and its environment in the nine-spin chain and in the square nine-spin cluster in the course of the ADRF on the basis of the criterion \cite{peres}. The PPT predicts the emergence of the spin pair entanglement at the same temperature as the Wootters criterion~\cite{wootters}. The PPT criterion uses the sum of the absolute values of negative eigenvalues after the partial transposition of the density matrix as the measure of entanglement. This measure is called the double negativity~\cite{vidal}. We use the criterions of entanglement~\cite{wootters,peres} for the investigation of the entangled states in a nine-spin chain and a plane nine-spin cluster in the next section. Our choice of nine-spin systems is limited by available computational power.
Notice that even best available supercomputers do not allow one to
calculate spin dynamics when the number of spins in the system
exceeds 16~\cite{guinz}. It is worth to noticing that we consider
spin-spin interactions of all spins in contrast to the approximation
of the nearest neighbor interactions when the number of spins in the
system can be significantly increased~\cite{lacelle}.
\section{Numerical analysis of entanglement in a nine-spin chain at
the ADRF} 
\begin{figure}[h] \begin{center} \includegraphics[width=8.6cm]{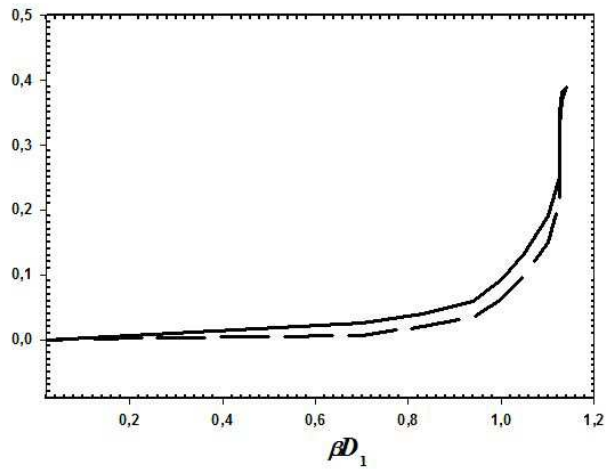}
\end{center}
\caption{The concurrence, $C,$ versus the dimensionless inverse
temperature, $\beta D_1,$ for spins 1 and 2 of the nine-spin chain
according to the Wootters criterion. The dash line shows the same
result obtained with the PPT criterion; $w_1/D_1=2$.} \label{fig1}
\end{figure}

The results of the numerical investigation of the spin
pair entanglement for a linear chain consisting of nine spins
coupled by the DDI at the ADRF are represented in figure~\ref{fig1}.
 We started the ADRF with the offset $\Delta_0=10D_1$
where $D_1$ is the DDI coupling constant of nearest neighbors in the
chain. The numerical calculations are performed for
$\Delta_n=(10-n)D_1,\ n=0,1,\ldots, 10.$ The corresponding values of
the inverse temperatures, $\beta_n=\hbar/kT_n\ (n=0, 1, \ldots,
10),$ are found from \eq{eentr} for the dimensionless entropy
$S/k=0.5.$ Figure~\ref{fig1} shows the concurrence of spins 1 and 2
of the nine-spin chain as a function of the dimensionless parameter
$\beta D_1.$ At comparatively high temperatures the concurrence is
equal to zero (see figure~\ref{fig1}) and the spin system is in a
separable state. When the temperature is getting sufficiently low in
the course of the ADRF the concurrence is sharply increasing and
entanglement emerges. The entangled state appears at $\beta
D_1\approx1.1$ (see figure~\ref{fig1} ), i. e. at $T\approx0.5\mu K$
when $D_1=2\pi10^4\ s^{-1}.$ Notice that ordered states of nuclear
spins were observed in a $CaF_2$ single crystal at microkelvin
temperatures \cite{abraham}. \begin{figure}[h] \begin{center}
\includegraphics[width=8.6cm]{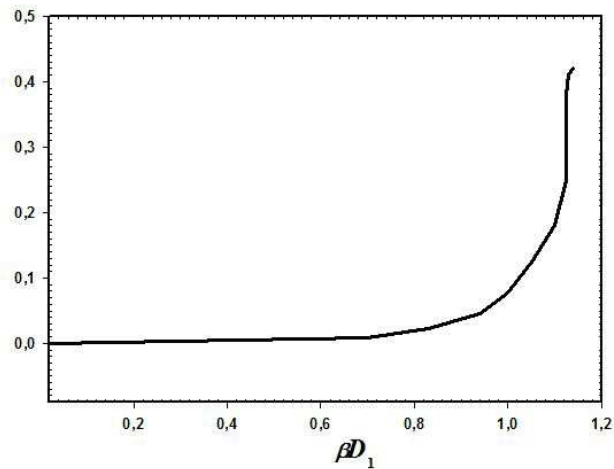}
\end{center}
\caption{The double negativity versus the dimensionless inverse
temperature, $\beta D_1,$ for spin 1 of the nine-spin chain and the
other spins; $w_1/D_1=2$.} \label{fig2} \end{figure}
\begin{figure}[h] \begin{center} \includegraphics[width=8.6cm]{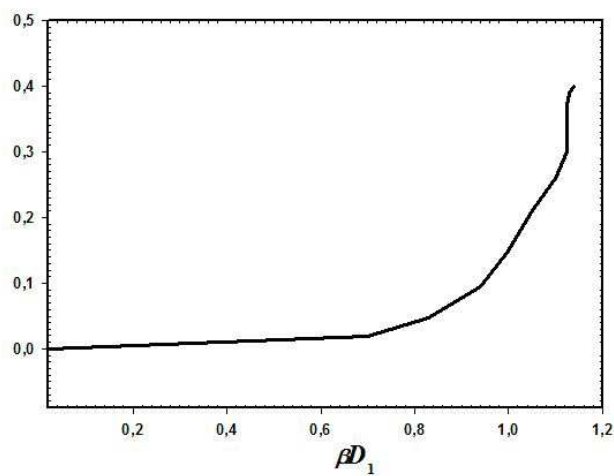}
\end{center}
\caption{The double negativity versus the dimensionless inverse
temperature, $\beta D_1,$ for the first three spins of the nine-spin
chain and the other spins; $w_1/D_1=2$.}\label{fig3} \end{figure}

The dash line in figure~\ref{fig1} demonstrates that the result for
the spin pair entanglement obtained with the PPT criterion coincide
practically with the ones obtained with the Wootters criterion.
Figure~\ref{fig2} shows the double negativity versus the
dimensionless parameter $\beta D_1$ for the first spin of the chain
(the first subsystem) and the other spins of the chain (the second
subsystem) in the course of the ADRF when the dimensionless entropy
$S/k=0.5.$ One can conclude that entanglement emerges at $\beta
D_1\approx1.1.$ This result is close to the one for the spin pair
entanglement. Figure~\ref{fig3} shows the double negativity versus
the parameter $\beta D_1$ for the first three spins of the chain
(the first subsystem) and the other six spins of the chain (the
second subsystem) at the same conditions. Here entanglement is
getting sufficiently large at $\beta D_1\approx1.1.$ In fact we have
found that entanglement of different subsystem emerges approximately
at the same temperature. \section{Entanglement in the square cluster
of nine spins } \begin{figure}[h] \begin{center}
\includegraphics[width=11.6cm]{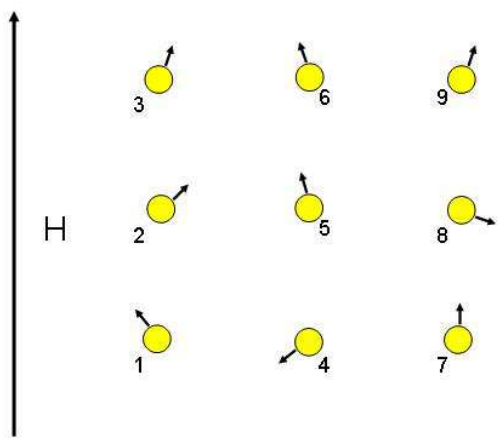}
\end{center}
\caption{The square cluster of nine spins.}\label{fig4} \end{figure}
The suggested method allows us to study entanglement in the systems
of arbitrary space dimensions. As an example, we consider the square
cluster consisting of nine spins (see figure~\ref{fig4}). The
dipolar coupling constant of spins j and k (the numbers of spins are
pointed in figure~\ref{fig4}) is \begin{equation}
D_{jk}=\frac{\gamma^2\hbar}{r_{jk}^3}(1-3\cos^2\theta_{jk}),
\end{equation}
where $\gamma$ is gyromagnetic ratio, $r_{jk}$ is the distance
between spins $j,\,k$ and $\theta_{jk}$ is the angle between the
vector, $\vec{r}_{jk},$ and the external magnetic field,
$\vec{H}_0.$ The simple analysis yields \begin{equation}
\cos^2\theta_{jk}=\frac{9(\{(j-1)/3\}-\{(k-1)/3\})^2}{([(j-1)/3]-[(k-1)/3])^2+9(\{(j-1)/3\}-\{(k-1)/3\})^2},
\end{equation}
and
\begin{equation} r_{jk}=a\sqrt{([(j-1)/3]-[(k-1)/3])^2+9(\{(j-1)/3\}-\{(k-1)/3\})^2},
\end{equation}
where $a$ is the distance between the nearest neighbors in the
cluster, $[q]$ is the integer part of $q$ and $\{q\}$ is the
fractional part of $q$. \begin{figure}[h] \begin{center}
\includegraphics[width=11.6cm]{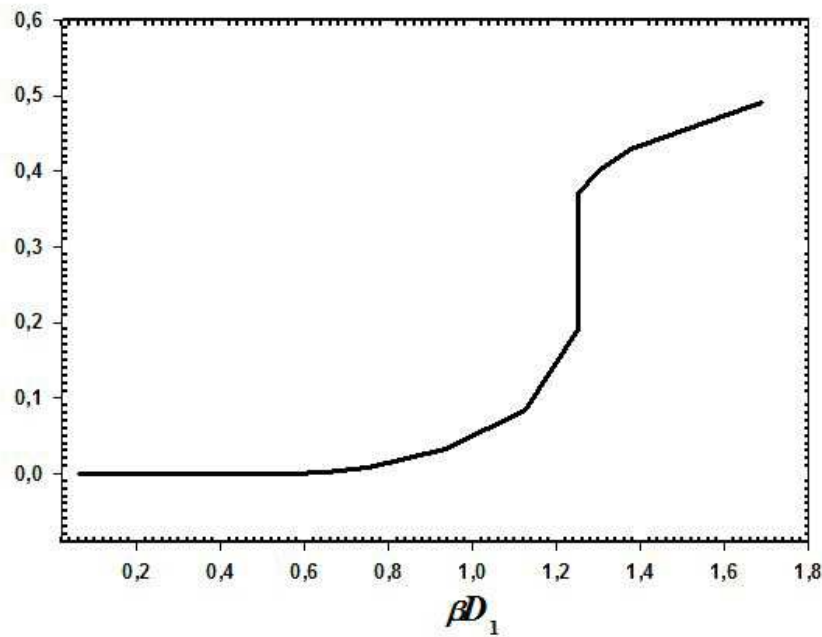}
\end{center}
\caption{The double negativity versus the dimensionless inverse
temperature, $\beta D_1,$ for spins 1, 2, 4 (first subsystem) and
spins 5, 6, 8, 9 (second subsystem), in the square cluster;
$w_1/a=2$.}\label{fig5} \end{figure}

\begin{figure}[h] \begin{center}
\includegraphics[width=11.6cm]{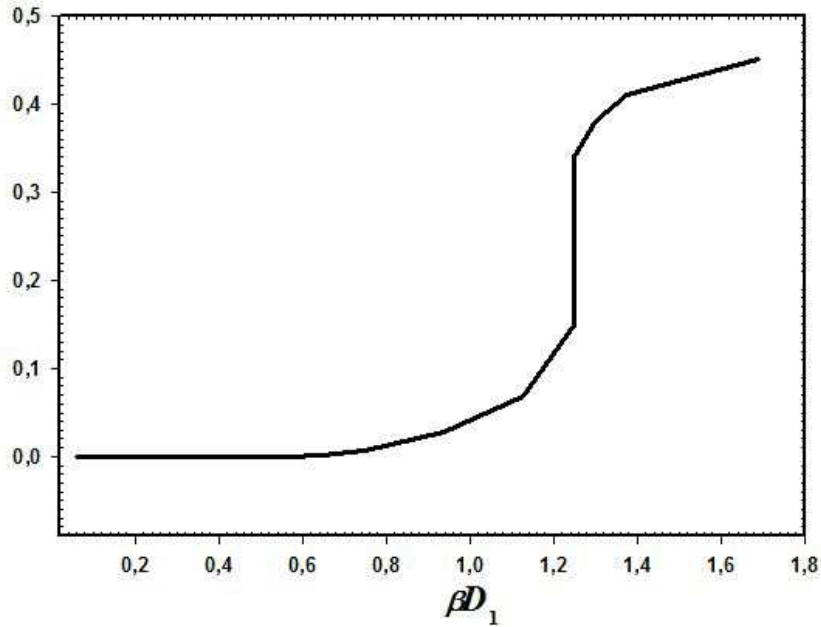}
\end{center}
\caption{The double negativity versus the dimensionless inverse
temperature, $\beta D_1,$ for spins  1, 2 (first subsystem) and
spins 3, 6, 9 (second subsystem), at $w_1/a=2$ in the square cluster
of nine spins.}\label{fig6} \end{figure}

Figure~\ref{fig5} and figure~\ref{fig6} show that the entangled
states emerge at $\beta D_1=1.3$ in the two-dimensional cluster.
Although the temperatures of the appearance of entanglement are
below here than in the one-dimensional case, they are almost the
same for different subsystems. The processes of destructive
interference are more effective in the two-dimensional cluster than
in the one-dimensional chain. They lead to a loss of many-spin
correlations which are responsible for the appearance of the
entangled states. Thus the temperature of the emergence of
entanglement is lower in the two-dimensional cluster than in the
one-dimensional chain.

\section{Conclusion}
We investigated numerically entanglement in the chain of nuclear
spins and in the square cluster with the DDI in the course of the
ADRF using a special computer program. We showed that the entangled
states emerge at microkelvin temperatures for typical DDI coupling
constants. Two criterions~\cite{wootters,peres} of entanglement
yield the same results for the spin pair entangled states.
Entanglement of different subsystems emerges approximately at the
same temperature and the pairwise entanglement can be used as an
indicator of entanglement of bigger subsystems. It is also worth to
notice that we take into account the DDI of the remote spins in
contrast to the works \cite{pyr, jetp} where the nearest-neighbor
interactions were only considered. The performed calculations show
that there are no the entangled states of remote spins both in
one-dimensional and two-dimensional cases. Entanglement emerges only
if subsystems are in a direct contact. Entanglement of different
subsystems at microkelvin temperatures suggest possible applications
of linear spin chains in quantum information processing.

We thank Professor D. E. Feldman and Professor V. A. Atsarkin for
stimulating discussions and E. I. Kuznetsova for assistance in our
work. This work was supported by the Russian Foundation for Basic
Research (grant 07-07-00048).

\end{document}